\documentclass[%
 aip,
 amsmath,amssymb,
 reprint,%
]{revtex4-1}

\usepackage{graphicx}
\usepackage{dcolumn}
\usepackage{bm}
\usepackage{xcolor}
\usepackage[utf8]{inputenc}
\usepackage[T1]{fontenc}
\usepackage{mathptmx}
\usepackage{etoolbox}

\makeatletter
\def\@email#1#2{%
 \endgroup
 \patchcmd{\titleblock@produce}
  {\frontmatter@RRAPformat}
  {\frontmatter@RRAPformat{\produce@RRAP{*#1\href{mailto:#2}{#2}}}\frontmatter@RRAPformat}
  {}{}
}%
\makeatother
\begin{document}

\preprint{AIP/123-QED}

\title[Automated classification of individual atoms on surfaces using machine learning]{Automated classification of individual atoms on surfaces using machine learning}

\author{Angéline Lafleur}
    \affiliation{Department of Physics, University of Ottawa, Ottawa, Ontario K1N 9A7, Canada}
    \affiliation{Center for Quantum Nanoscience, Institute for Basic Science (IBS), Seoul, 03760, Republic of Korea\\}%
    \affiliation{Department of Physics, Ewha Womans University, Seoul, 03760, Republic of Korea}
    
\author{Soo-hyon Phark}%
    \affiliation{Center for Quantum Nanoscience, Institute for Basic Science (IBS), Seoul, 03760, Republic of Korea\\}%
    \affiliation{Department of Physics, Ewha Womans University, Seoul, 03760, Republic of Korea}
    \email{phark@qns.science}

\date{\today}

\begin{abstract}
Leveraging scanning tunneling microscopy (STM) for atomic-scale fabrication has led to many advancements such as the creation of atomic electron-spin qubit structures on surfaces. However, the time-consuming and tedious nature of this process calls for improvements, and this study explores the use of machine learning (ML) to automate certain steps, notably identifying appropriate atomic candidates for the structures. We classify titanium and iron atoms on a magnesium oxide (MgO) surface, which are prototypical on-surface spin qubit candidates, showing distinct topographic and spectroscopic features depending on the bonding sites of the MgO surface. Employing a semi-automated computer vision process, we train a convolutional neural network with STM topographic images and scanning tunneling spectroscopy (STS) curves of several hundred atoms. After training, the topography model achieves an 86\% validation accuracy in classifying 200 new images, and the STS model, a 100\% accuracy for a sample size of 100 atoms. This method extends its applicability to various nanoscopic measurements, including atomically resolved imaging and local probing of electronic properties, offering a promising approach for classifying atoms and molecules on surfaces.
\end{abstract}

\maketitle

\section{\label{sec:level1}Introduction}
Scanning probe microscopy (SPM) allows for detailed atomic-scale characterization of topographic, electrical, and magnetic properties of materials, as well as molecules and atoms on surfaces. SPM, particularly Scanning Tunneling Microscopy (STM), has also led to atomic-scale fabrication technology. Using the STM tip, nanostructures can be built atom-by-atom \cite{eigler_positioning_1990, oura_atomic_2003, wyrick_enhanced_2022}, allowing for studying chemical reactions with individual molecules \cite{matthiesen_observation_2009}, or building atomic chains \cite{hla_atom-by-atom_2014}, molecular cascades \cite{Heinrich2002, hla_atom-by-atom_2014}, spintronic logic devices \cite{Khajetoorians2011,hla_atom-by-atom_2014}, or single atom transistors \cite{Fuechsle2012, lu_high-throughput_2022}. Notably, combining Electron Spin Resonance (ESR) with STM~\cite{Baumann_science_2015} allows the creation of quantum processors using atoms on surfaces. Universal quantum control and multi-qubit operations were achieved using titanium (Ti) atoms adsorbed on an ultrathin magnesium oxide (MgO) layer, demonstrating the viability of this platform for investigating spins for quantum information.~\cite{wang_universal_2023,phark_advsci_2023,Wang2023}

When adsorbed on a solid surface, single atomic species often show distinct charge and/or spin configurations depending on the environments, such as bonding sites, nearby defects, and thickness of surface layer. Therefore, the first step of the above-mentioned quantum technology based on on-surface single spins is to characterize individual atoms and classify them into each adsorbate group of identical physical and chemical properties. However, this step has been performed mostly by manual microscopy and spectroscopy of atoms one-by-one due to the inherent shortcomings of the STM as a real space probe, such as variance in STM/S measurements stemming from uncontrollable tip-apex and time-dependent spacial drift of the probing tip position. This has rendered experiments using a STM to be very time consuming against cost of sub-atomic scale measurement precision, which still make an automation of STM-based quantum technology a challenging task.

With the rising popularity of Artificial Intelligence (AI), there have been advances in SPM automation employing various ML methods. \cite{azuri_role_2021,Ziatdinov2017,Ziatdinov2022, kurban_atom_2021, Krull2020,Rashidi2018, ziatdinov_learning_2017, wang_machine_2020} Particularly, various studies explore the creation of STM robots which use automated programs to operate the STM controls and ML to characterize the sample and tip, creating a feedback system. \cite{Krull2020,Rashidi2018} 

This work presents an automation of the identification and classification of adsorbed atoms on a surface for the creation of nanostructures. Two ML models are investigated: one which classifies atoms based on their topography profile, and the other based on their STS spectra. Our method can accelerate steps such as finding atomic candidates of spin qubits and magnetic atoms to build a spin-polarized tip~\cite{Yu_2023} for spin-polarized STM in general,~\cite{Yamada2003} e.g., driving and probing quantum coherent control of spins on a surface. The models are trained and tested on titanium (Ti) and iron (Fe) atoms adsorbed on an MgO surface, physical systems used to make atomic scale qubit structures.~\cite{Wang2023} The models classify detected atoms into one of four categories: Ti adsorbed on an oxygen site, Ti adsorbed on a bridge site, Fe adsorbed on an oxygen site, and a category grouping all other unknown types. As more complex multi-qubit structures are built and various atomic qubits are considered, the number of classes increases, and the automated classification process will help further accelerate the fabrication of quantum processors based on ESR-STM.

\section{\label{sec:level1}Methodology}
\subsection{\label{sec:level2}Data acquisition}

\begin{figure*}
\includegraphics[]{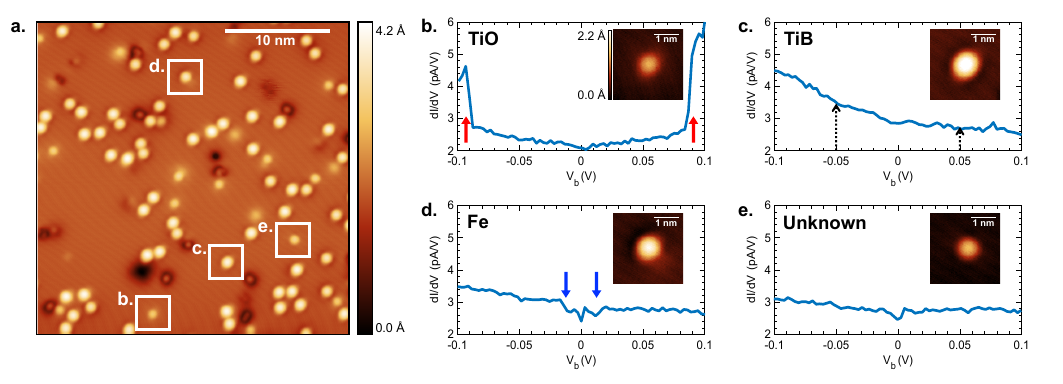}
\caption{\label{fig:topo} STM topography and STS data of sample surface. (a) STM image ($30 \times 30$ nm$^2$) of Ti and Fe adsorbed on MgO, measured at $V_{\rm b}$ = 100 mV and $I_{\rm set}$ = 10 pA. The Ti atoms adsorbed at oxigen (TiO) and bridge (TiB) sites, iron (Fe) atoms, and unknown atoms characterized in (b-e) are highlighted. (b-e) STS spectra of the 4 atomic species with their corresponding topography images in the insets ($3 \times 3$ nm$^2$): (b) TiO, (c) TiB, (d) Fe, and (e) an unidentified defect (Unknown).
}
\end{figure*}

The ML model was trained on STM topography and spectroscopy data of Ti and Fe adbsorbed on two-monolayer MgO patches formed on Ag(100). The sample was fabricated by first preparing an atomically clean Ag surface by Ar ion sputtering and annealing cycles. An MgO layer was then grown on the Ag surface by evaporating Mg in an $\text{O}_2$ atmosphere of $1.0\times10^{-6}$ mbar at 580 K. Fe and Ti atoms were deposited on the pre-cooled MgO surface. All measurements were performed in a commercial STM (USM1300, Unisoku) operating at 0.4 K, with mechanically cut Pt/Ir tips. The tip was poked into the Ag(001) surface until satisfactory observation of topographic and spectroscopic features of the atomic adsorbates~\cite{Baumann_prl_2015,Yang_prl_2017} was achieved. To broaden usage of the model, training data was taken with several different tips, leading to variations in some topography images caused by the specific tip-apex structure of each tip.

STM measurements were conducted in constant current mode with setpoint current ($I_{\rm set}$) of 10 pA and bias voltage ($V_{\rm b}$) of 100 mV applied to the sample, as a topographic image is shown in Fig. 1(a). The resolution of each scan was 28.8 pixels per nm. This prevents artifacts originating from resizing the image from influencing the ML model, which takes as input $86 \times 86$ pixel images. Topography data was acquired in various tip conditions to create a tip condition-insensitive model, up to a certain tolerance. Scans were taken with various sizes and at different locations on the MgO surface, in areas with minimal defects (hydrogen, etc.). In total, topography data of 1000 individual atoms was collected and used to train and analyze the ML model.

Each atom was probed using STS to determine its electronic features, TiO – titanium bound to an oxygen site, TiB – titanium bound to a bridge site, Fe – iron bound to an oxygen site, and unknown – defects or atoms with abnormal tunneling spectra. Characteristic spectra for each of these categories are found in Figs. 1(b)-(e).

In all STS measurements $V_{\rm b}$ was swept from $-$100 mV to 100 mV with 71 differential conductance ($dI/dV$) data points.
Spectra on TiO and Fe show characteristic features, as indicated by the red and blue arrows, respectively, stemming from inelastic tunneling of electrons.~\cite{Yang_prl_2017,Baumann_prl_2015} 
No specific spectral feature has been reported for TiB in the $\lvert V_{\rm b}\rvert$ range $\leq$ 100 mV. However, a $\sim$50 \% decrease in differential conductance was reproducibly observed across the zero bias,~\cite{hwang_rsi_2022} as indicated by black dotted arrows, while contribution of direct tunneling between tip and substrate is $V_{\rm b}$-independent and featureless. 
Once an atom is classified, it is labeled according to its position in the larger STM scan. An array of images of atoms as well as their labels is used to train the ML model.

\subsection{\label{sec:level2}Image pre-processing}
To maximize the performance of the ML model, it is crucial to isolate the atom of interest in the image, and minimize differences other than the atomic signatures. Therefore, each image is zoomed in and centered on each atom of interest using the same size (3 $\times$ 3 nm) and resolution (86 $\times$ 86 pixels). To accomplish this efficiently from a large STM scan containing 50 to 100 atoms, a semi-automated processing stage is implemented using computer vision, as shown in Fig. 2. The challenge lies in the fact that for all training images, the correct label for each atom must be preserved. The developed Python script successfully takes as input large topography images and generates the training datasets by putting the images in the folders corresponding to each of their classes.

The program sequentially processes each imported scan file, allowing for a larger scale automation. To import the STM topography data stored in a .sxm file generated from the STM controller (Nanonis, SPECS), the pySPM library is used.~\cite{pyspm} For each scan file, the program first extracts the topography data (Z-channel), and then checks the scan direction and performs necessary corrections to ensure proper image orientation. Plane subtraction is applied through pySPM to correct for any tilt in the substrate surface, producing a corrected image saved as a temporary file.

Next, using the OpenCV (cv2) package for Python~\cite{opencv_library}, computer vision processing steps are implemented to isolate the position of each atom. The image is first converted to grayscale with the highest topography valued pixel set to white, and lowest set to black. This essentially encodes the height values with 8 bits resolution, normalized across atoms from the image. A Gaussian blur is applied to smooth the image, followed by binary thresholding to create a binary image highlighting regions of interest, which are potential atoms. The threshold is set to half of the total levels, then every pixel higher than the threshold will be set to white, and all other pixels to black. Since all the atoms are higher than the background, this results in the atoms showing up as white spots. Contour detection can then be used to locate each of these spots. For each detected contour, several checks are performed to ensure it represents a single atom. They are filtered based on their perimeter to exclude excessively large regions representing high points of the substrate material or other artifacts from the threshold level selected. Then by calculating the positions of the valid contours' centers, cropped images can then be generated around the center of each atom in the original unaltered image. 

\begin{figure}
\includegraphics[]{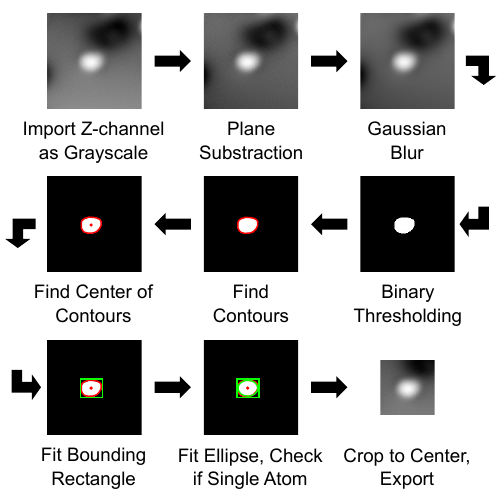}
\caption{Workflow overview of image pre-processing steps to automatically extract image of an individual atom from larger STM scans.
}
\end{figure}

However, when atoms are close to each other, they are often not distinguishable in the binary image, showing up as a single contour. Most commonly, two atoms next to each other would have a single "peanut-shaped" contour. To identify these, a bounding box surrounding the contour is created, and an ellipse is fitted to the dimensions of that bounding rectangle. If the contour contains a single atom, it will have a mostly circular shape, so an ellipse fitted inside of its bounding rectangle should have an area quite similar to it. On the contrary, if the contour contains two atoms, the area of the fitted ellipse would be significantly larger than the "peanut" contour's area. When a "peanut" contour is detected, the region undergoes a second round of thresholding, raising the threshold closer to the apexes of the atoms' topographies such that the atoms can be differentiated. All previously mentioned steps are repeated to verify that the new contours really represent single atoms, if not, the thresholding is repeated a third time at a higher level. Once each contour is verified to contain a single atom, the contour's center point position is found, and a cropped area of 3 $\times$ 3 nm$^2$ centered on that point in the original image is saved.

The above process is used for training data for the model, as well as new atom images to be classified by the model. For training data, each saved image must be classified in its corresponding folder (TiB, TiO, Fe, Unknown) to be processed into the model. When acquiring STS data for each atom in the STM scan, each atom is identified by the user. To transfer this information to the Python program, a Comma-separated Values (CSV) file listing the type names in order is created. The OpenCV contours detection function outputs a list of each contour in order from its position in the image starting from the bottom left corner. A separate program with contour detection was created to print the number in which each atom's contour is detected on the image. By following the numbering order of that image, the user can create the CSV list of classes for each of those atoms by number. With the CSV file loaded into the pre-processing program, it can then save each cropped image into the folder with the name dictated by the element in the CSV file corresponding to the number of that atom.

Once this program is run for all the desired STM scan images, the training dataset can be used to create the ML model. The subsequent model was trained on 995 images with 248 Fe, 475 TiB, 133 TiO, and 139 unknown, reflecting their presence on the sample.

\begin{figure*}
\includegraphics[]{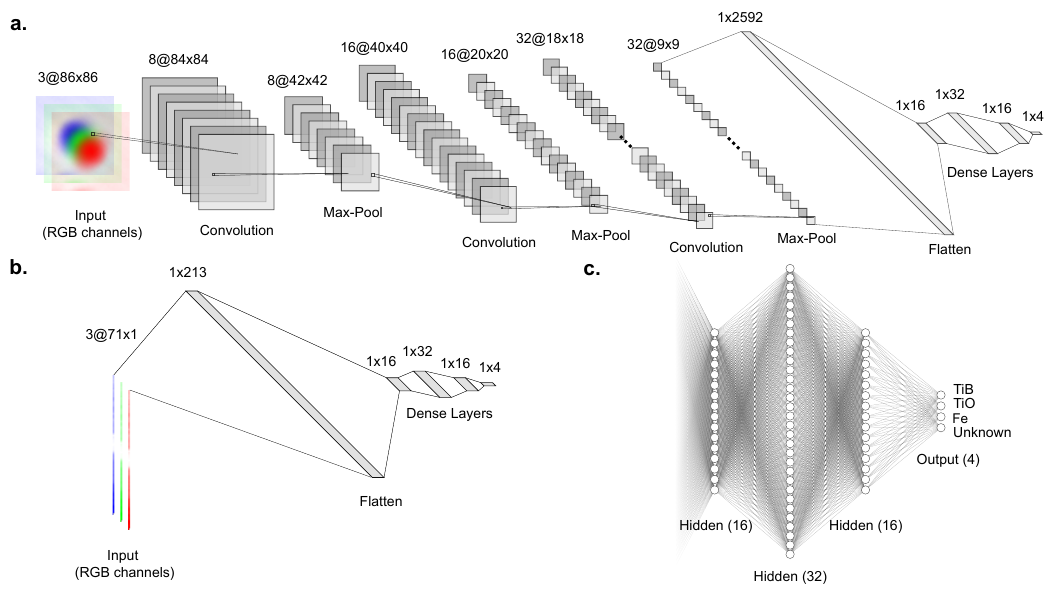}
\caption{\label{fig:model} ML model visualization. (a) Schematic of Convolutional Neural Network (CNN) used for STM image classification showing convolution, max-pooling, and dense layers. (b) Schematic of Feedfoward Neural Network (FNN) with normalizing layer to process STS curves as 1D images. (c) Focused view of the dense layers used for classification in (a-b).
The NN-SVG package was used to facilitate the creation of the schematics.\cite{lenail_nn-svg_2019}
}
\end{figure*}

\subsection{\label{sec:level2}Topography model: Convolutional Neural Network}
Convolutional Neural Networks (CNN) have been shown to accurately classify images of fixed size and resolution. \cite{chen_review_2021} To classify the images of atoms obtained in the pre-processing steps, a CNN model was created in Python, using the TensorFlow and Keras libraries. \cite{TF,chollet2015keras} It is composed of layers that progressively extract features from the images, and then classify the atoms based on those features. This is done by alternating convolution and pooling layers for feature identification, followed by fully-connected layers for classification, as shown in Fig. 3(a). 

In TensorFlow, the input layer of the model consists of the extracted  red, green, and blue (RGB) channels of the image. In this case, the image of the atom  is encoded from a single Z-channel, so the RGB channels all contain the same data. Therefore, the input layer contains three identical arrays of $86 \times 86$, corresponding to the dimensions of the image. Each value in the array encodes the topography at that pixel, and is proportional to the pixel brightness in the image.

The first convolutional layer applies 8 filters with a kernel size, or window, of $3 \times 3$ to detect simple features in the image. To capture an increasing number of details, subsequent convolutional layers of more filters (16 and 32) are applied. The model uses the rectified linear unit (ReLu) activation function to calculate the output nodes from their inputs and weights in these layers. Following each convolutional layer, a max-pooling layer with pool size, or window, of $2 \times 2$ is used to reduce the image size while preserving its characteristics.

Next, the feature maps are flattened and passed through dense layers, see Fig. 3(c), reducing the system to a simple neural network with 2502 input nodes and 4 output nodes that produce probabilistic classifications using softmax activation. To further prevent overfitting, the model randomly drops out 20\% of input values of the flattening layer, 5\% after the first 16 neuron hidden layer, and another 5\% after the 32 neuron layer.

To train the model, first, the data is loaded, augmented, randomly shuffled, and divided into training and validation sets with a validation split of 20\%. Then the model is trained using the Adam optimizer, a stochastic gradient descent method based on the adaptive estimation of first and second order moments.~\cite{kingma_adam_2017} The loss function chosen to be minimized in the gradient descent is the sparse categorical cross-entropy loss.~\cite{10351209} It is defined as the difference between the true probability distribution of the output classes and the probability distribution predicted by the model. 

\subsection{\label{sec:level2}STS model: Feedforward Neural Network}
To classify the atoms based on their STS data, since the data is one-dimensional, no convolution layers are required, and a Feedforward Neural Network (FNN) is adequate.  \cite{329294} However, to simplify the design and take advantage of the TensorFlow libraries, the STS spectrum is converted into an image. All the $dI/dV$ versus bias voltage curves span the range from $-100$ mV to 100 mV in 71 points. With the same x-axis for each curve, the $dI/dV$ data points can instead be represented as a 1D array. This array is converted to an image of $71 \times 1$ pixels, where the $dI/dV$ value is mapped to a gray scale brightness value. These images are then fed into the following ML model, see Fig. 3(b).

The TensorFlow package interprets the images as RGB 3-channel inputs which then get flattened to a 1D array of neurons. Next, dense layers identical to the ones in the topography model are used to extract features and classify the spectra, see Fig. 3(c). This structure is optimal to extract and identify the features specific for the TiO spectra, as well as the more subtle differences found in the Fe, TiB, and unknown spectra.

Before training the model, the dataset is normalized to values between 0 and 1, then shuffled, and divided into training and validation sets with an 80-20 split. The model is then compiled using the Adam optimizer using the Sparse Categorical Crossentropy loss function. No dropout layers are needed because the model does not tend to overfit since the STS data is uniform, unlike the topography images which contain scanning artifacts and are more complex.

\section{\label{sec:level1}Results and Discussion}
\subsection{\label{sec:level2}Topography model results}
The CNN model was trained and validated on 995 classified images of atoms and recorded its performance using two metrics: the sparse categorical crossentropy loss, as well as the accuracy in percentage. A peak accuracy was achieved After 200 epochs of training, the validation accuracy tends to 86\%, with a peak accuracy achieved of 87.9\% at the $\text{196}^{\text{th}}$ epoch indicating the model's performance on new data. This corresponds to a minimum in the loss function of 0.41, see Fig. 4. It can be noted that at the beginning of training, the validation has a lower accuracy and higher loss, and for higher epochs, the validation loss plateaus, as the training loss continues to drop. The drop-out layers which slow down the fitting, cause the training data to perform inferior to the validation set for early epochs, as some knowledge of the features is dropped. But as the model gains knowledge of the features of the training data at higher epochs, the training loss function decreases. If the model is trained longer than 200 epochs, this overfitting takes over, and the model no longer has similar training and validation response.

The peak accuracy achieved indicates a useful knowledge gained of the features of the atoms' topographies, and provides an estimate of the atoms' classes. However, an automated classification process based only on this model, would be not enough to be entirely relied upon.

\begin{figure}
\includegraphics[]{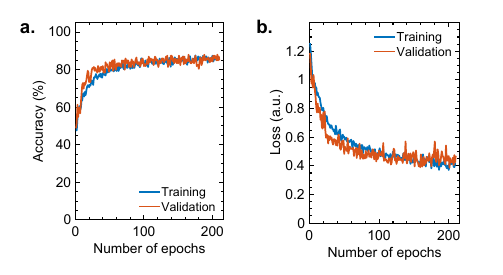}
\caption{Results of the STM topography model. (a) Percentage of correct classifications and (b) sparse categorical crossentropy loss of the training and validation sets as a function of number of training cycles.
The PlotPub package was used to create these plots.\cite{Habib2024}
}
\end{figure}

\subsection{\label{sec:level2}STS model results}
The STS spectra FNN model was trained on less datapoints than the CNN, 494 images are used. The training performance over 20 epochs can be seen in Fig. 5. The model rapidly learns the necessary features to correctly classify the atoms by their STS spectra. The model reaches a 100\% training and validation accuracy over the sample size, and reaches a minimum loss of $2.7 \times 10^{-4}$.

With such high performance, this model is entirely adequate for the rapid and accurate classification of TiO, TiB, Fe, and unknown atoms. The model converges fast and faces no overfitting problems, even with a smaller training dataset compared to the topography model.

\begin{figure}
\includegraphics[]{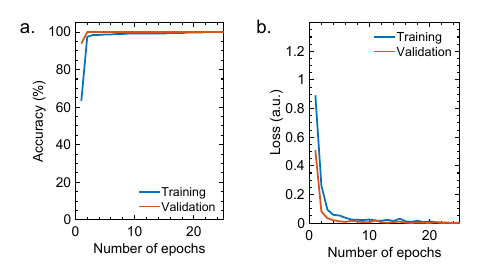}
\caption{Results of the STS model. (a) Percentage of correct classifications and (b) sparse categorical crossentropy loss of the training and validation sets as a function of number of training cycles.
The PlotPub package was used to create these plots.\cite{Habib2024}
}
\end{figure}

\section{\label{sec:level1}Conclusion}
This work provides a summary of the best results achieved when using TensorFlow ML models to classify single Ti and Fe atoms on a MgO surface. It shows promise as an approach to automate the search for types of atoms on surfaces and could be used in a fully automated STM measurement setup. The near perfect accuracy obtained renders this spectroscopy-based ML model ideal for classifying atoms. When paired with the computer vision pre-processing steps mentioned to locate each atom, a fully automated large scale atom classification can be achieved. Further work is required to achieve the precision needed for topography-based atom classification to replace other methods in a fully automatized scheme. Notably, for the model to be robust against artifacts from tip variations, a larger training dataset obtained from diverse tip conditions could be used to increase accuracy. Furthermore, more specific classes, instead of grouping all defects in an unknown category would help for more accurate feature learning. This would allow for a faster and simpler classification process than with the STS model only, allowing for applications in identifying various STM features extending beyond single adsorbed atoms.

\begin{acknowledgments}
AL acknowledges financial support from the University of Ottawa (Faculty of Engineering International Experience Bursary).
SP acknowledges financial support from the Institute for Basic
Science (IBS-R027-D1).

\end{acknowledgments}

\bibliography{Atom_Classification}

\end{document}